\documentclass[aps,amsmath,amssymb,twocolumn,superscriptaddress]{revtex4}
\usepackage[]{graphicx}
\usepackage{color}
\usepackage{bbold}
\usepackage{braket}
\usepackage[colorlinks=true, pdfstartview=FitV, linkcolor=blue, citecolor=red, urlcolor=black]{hyperref}
\newcommand{\be}{\begin{equation}}
\newcommand{\ee}{\end{equation}}
\newcommand{\ben}{\begin{eqnarray}}
\newcommand{\een}{\end{eqnarray}}
\newcommand{\bes}{\begin{subequations}}
\newcommand{\ees}{\end{subequations}}
\def\bal#1\eal{\begin{align}#1\end{align}}
\usepackage{comment}

\newcommand{\bfi}{\begin{figure}}
\newcommand{\efi}{\end{figure}}
\newcommand{\bc}{\begin{center}}
\newcommand{\ec}{\end{center}}
\newcommand{\sech}{\mbox{sech}}

\begin{document}
    \title{Cuscuta-Galileon Braneworlds}
    \author{D. Bazeia}
    \affiliation{Departamento de F\'\i sica, Universidade Federal da Para\'\i ba, 58051-970 Jo\~ao Pessoa, PB, Brazil}
    \author{A.S. Lob\~ao Jr.}
    \affiliation{Escola T\'ecnica de Sa\'ude de Cajazeiras, Universidade Federal de Campina Grande, 58900-000 Cajazeiras, PB, Brazil}
     \author{M.A. Marques}
    \affiliation{Departamento de Biotecnologia, Universidade Federal da Para\'iba, 58051-900 Jo\~ao Pessoa, PB, Brazil}
    \author{R. Menezes}
    \affiliation{Departamento de Ci\^encias Exatas, Universidade Federal da Para\'iba, 58297-000 Rio Tinto, PB, Brazil}

\begin{abstract}
We investigate braneworlds modeled by topological solutions that arise from the so-called Cuscuta-Galileon model. We develop a first order framework and illustrate our procedure with the scalar field having the well-known hyperbolic tangent profile. We find conditions that must be imposed to the parameters of the model in order to have solutions connecting minima of the potential, with the brane constrained to interpolate Minkowski and anti de Sitter geometries. We also find solutions where the brane only interpolates anti de Sitter geometry. In both cases, the gravity sector of the brane is stable against small fluctuations of the metric. \end{abstract}
\maketitle

Braneworld models  arise in theories of gravity in $(4,1)$ spacetime dimensions. It was conceived in 1999 as a tentative to explain the hierarchy problem \cite{Randall:1999vf}. The original model supports the so-called thin brane, with the derivative of the  warp factor having a discontinuity at its center. By including scalar fields in the action, it was shown in Ref.~\cite{Goldberger:1999uk,Skenderis:1999mm,DeWolfe:1999cp,Csaki:2000fc} that a kinklike solution can give rise to thick branes. In the presence of scalar fields, it is known that modifications in the dynamics may generate interesting changes in the brane profile. In this, direction, several studies have addressed this issue over the years; see, e.g., Refs.  \cite{mirjam,Bazeia:2008tj,Adam:2007ag,Bazeia:2008zx,Liu:2009ega,Bazeia:2013euc} and references therein. 
Among the many possibilities in the study of braneworlds, one may find the presence of asymmetric structures; see, for instance, Refs.~\cite{mirjam2,as0,as1,as2,as3,as4,as5,as6,as7,as8,as9,as10}. The asymmetry can appear in different ways, for instance, one may have it as a consequence of interpolation of distinct geometries (see Ref.~\cite{mirjam2}). It may also be originated from the internal structure of the scalar field \cite{as9} or due to the asymptotic behavior of the solutions \cite{as10}.

Braneworlds may also be investigated in noncanonical models, in which the Lagrange density associated to the scalar field is a general function of the field and the kinetic term depending on its first derivative \cite{genbrane,genbrane2,genbrane3,genbrane4}. A particular model that has been gaining attention is the Cuscuton one, introduced in Ref.~\cite{asfshordi1,asfshordi2} in the context of cosmology. Since then, several papers dealing with the Cuscuton term have appeared in the literature; see Refs.~\cite{cusc1,cusc2,cusc3,cusc4,cusc5,cusc6,cusc7,cusc8}.
We can also modify the dynamics by including second order derivatives in the fields in the form $\nabla_a\nabla_b\phi$. This prescription is generally known as Horndeski theories or generalized Galileon theories \cite{Horndeski:1974wa,Nicolis:2004qq,Nicolis:2008in,Deffayet:2009wt,Deffayet:2010qz,DeFelice:2011aa}. In general, these theories obey the symmetry $\phi\to\phi+b_\mu x^\mu+c$, where $b_\mu$ is the constant vector and $c$ is a real number. Horndeski theories have been much investigated as they have led to new and distinguishable inflationary predictions; see, for example \cite{Gao:2011qe,DeFelice:2011uc,Kobayashi:2011nu,Clifton:2011jh,Fu:2019xtx}.

In recent studies, it was also considered the inclusion of both kinematic modifications presented above, with kinetic terms depending on the first and the second derivatives of the field. This is the basis behind the so-called Cuscuta-Galileon model, where both the Cuscuton and Galileon-like terms are included in the action simultaneously \cite{deRham:2016ged, Baker:2017hug,Panpanich:2021lsd,Maeda:2022ozc}. Horndeski theories have been relatively successful in describing aspects of the accelerated expansion of the Universe, and the recent results of Refs. \cite{Panpanich:2021lsd,Maeda:2022ozc} which nicely provide interesting sequence of the thermal cosmological history, have motivated us to investigate the possibility to construct braneworld model based on the Cuscuta-Galileon dynamics.

We start the present investigation by considering an action that describes a thick brane model in five dimensions of the spacetime sourced by a single real scalar field in the form
\begin{equation}\label{eq1}
    S=\!\int\! d^5x\sqrt{|g|}\left(-\frac14R+{\cal L}_s(\phi,X)\right),
\end{equation}
where $g$ is the determinant of the metric, $R$ is the Ricci scalar, ${\cal L}_s(\phi,X)$ is the Lagrange density and $X\equiv(1/2)\nabla_a\phi\nabla^a\phi$ represents the dynamical term associated to scalar field $\phi$. In this paper, Latin indexes $a,b,c$ run from $0$ to $4$ and Greek indexes $\mu,\nu$ run from $0$ to $3$.

The simplest Lagrange density that support stable braneworld configuration is $\mathcal{L}_s = X-V(\phi)$. In this case, the brane can be modeled by a kinklike solution which connects the minima of the potential $V(\phi)$ \cite{DeWolfe:1999cp}. As we have commented before, one may consider the inclusion of the Cuscuton term \cite{asfshordi1,asfshordi2}, with
\be\label{cuscuton}
\mathcal{L}_s = X + f(\phi)\frac{2X}{\sqrt{|2X|}}-V(\phi).
\ee
This was first investigated in Ref.~\cite{cusc7}. There, to keep the minima of the potential connected by the solution, it was considered a function $f(\phi)$ that goes to zero in the asymptotic limits of the solution. In this paper, we take a novel approach: inspired by the Cuscuta-Galileon model investigated in Ref.~\cite{Panpanich:2021lsd,Maeda:2022ozc}, we consider the inclusion of a Horndeski-like term in Eq.~\eqref{cuscuton} and take $f(\phi)=\beta$, where $\beta$ is constant. The new Lagrange density has the form
\begin{equation}\label{lagrange}
{\cal L}_s=X+\beta \frac{2X}{\sqrt{|2X|}}+\alpha\ln(|X|)\Box\phi - V(\phi).
\end{equation}
In this expression $\Box\equiv\nabla_a\nabla^a$, so our model has now the dynamics depending on a second derivative of the field. By varying the action with respect to the metric we get the equation
\be\label{einstein}
	 R_{ab}-\frac12g_{ab}R = 2T_{ab},
\ee
where the energy momentum tensor is given by
\be
\begin{aligned}
	T_{ab}&=\Big(1+\frac{\beta}{\sqrt{|2X|}}\Big)\nabla_a\phi\nabla_b\phi+\frac{\alpha}{X}\Big(\Box\phi\nabla_a\phi\nabla_b\phi \\
	&-\nabla_a X\nabla_b\phi-\nabla_b X\nabla_a\phi\Big) -g_{ab}\Big(X+\beta\, \frac{2X}{\sqrt{|2X|}}\\
	&-V-\frac{\alpha}{X}\nabla_c X\nabla^c\phi\Big).
\end{aligned}
\ee
The equation of motion that arises from the variation of the action with respect to the scalar field is
\be\label{field}
\begin{aligned}
	&-\nabla_a\Bigg(\left(1+\frac{\beta}{\sqrt{|2X|}}\right)\nabla^a\phi\Bigg)-\nabla_a\Big(\frac{\alpha}{X}\Box\phi\nabla^a\phi\Big)\\
	&+\nabla_a \Big(\frac{\alpha}{X}\nabla^aX\Big)=V_\phi.
\end{aligned}
\ee
By using the above equation, one can show that the energy momentum tensor is conserved, i.e., $\nabla_a T^{ab}=0$. Since we are interested to study braneworld scenario, we consider the line element in the form
\begin{equation}\label{ds2}
    ds^2=e^{2A}\eta_{\mu\nu}dx^\mu dx^\nu-dy^2\,,
\end{equation}
where $A$ is the warp function, $\eta_{\mu\nu}$ is the four-dimensional Minkowski metric with signature $(+,-,-,-)$ and $y$ is the extra dimension. In order to obtain localized solutions we consider static configurations assuming that the warp function and the scalar field only depend on the extra dimension $y$, so that $A=A(y)$ and $\phi=\phi(y)$. This makes  $X=-\phi'^2/2$ such that Eq.~\eqref{field} become
\be\label{eqmovphi}
\phi''+4A'\phi'+4\beta A'\text{sgn}(\phi')+8\alpha\big(A''+4A'^2\big)=V_\phi.
\ee
Here, $V_\phi=d V/d\phi$ and the prime stands for the derivative with respect to the extra dimension, i.e., $\phi'=d\phi/dy$, $A'=dA/dy$, etc. It is interesting to note that the above equation is of second-order, as in the usual Galileon model. Regarding the Einstein equation \eqref{einstein}, only two components survive, leading to
\bes\label{eqeinstein}
\bal
&3 A''=-2 \phi'^2-2 \beta |\phi'|+4 \alpha  \left(\phi ''-4 A'\phi'\right),\label{eqeinsteina}\\
&\frac{1}{2} \phi'^2-3{A'}^2+8 \alpha A' \phi'=V.\label{eqeinsteinv}
\eal\ees
It is possible to show that from the three equations \eqref{eqmovphi} and \eqref{eqeinstein}, only two are independent. We can then deal with the system of equations \eqref{eqeinsteina} and \eqref{eqeinsteinv}, noticing that these equations are invariant under the change $y\to-y$. For simplicity, we consider only monotonically increasing solutions for $\phi$, having a kinklike profile. 

We then investigate how the brane behaves for solutions with exponential tails. To do so, we consider that, at $y\to\pm\infty$, the scalar field obeys 
\be\label{phiasy}
\phi-v_\pm= \kappa_{\pm}\, e^{- m_\pm |y|},
\ee
where $v_\pm$ denotes the asymptotic values of the solution, $\phi(y)$, at $y\to\pm\infty$, and $\kappa_\pm$ and $m_\pm>0$ are constants which depend on the specific model under investigation. In the standard case ($\alpha=\beta=0$), Eqs.~\eqref{eqeinstein} read
\bes
\bal \label{allstandard}
& 3 A''=-2 \phi'^2,\\ \label{vstandard}
& V=\frac12{\phi'}^2-3A'^2.
\eal
\ees
Substituting Eq.~\eqref{phiasy} in Eq.~\eqref{allstandard}, we get
\be
A^\prime = {A^\prime}_\pm \pm \frac{m_\pm\kappa_\pm^2}{3} \,e^{- 2m_\pm |y|},
\ee
in which ${A^\prime}_+$ and ${A^\prime}_-$ are both constants of integration that represent the asymptotic values of the derivative of the warp function at $y\to\pm\infty$. So, if ${A^\prime}_+=0$ (${A^\prime}_-=0$) the brane engender a Minkowski geometry at $y\to+\infty$ ($y\to-\infty$). On the other hand, if ${A^\prime}_+<0$ (${A^\prime}_->0$), we have an anti de Sitter $(AdS)$ geometry. To find how the potential behaves, we can use Eq.~\eqref{vstandard} to get $V(v_\pm) = -3{A^\prime}_\pm^2$, $V_\phi(v_\pm) =0$ and
\be
V_{\phi\phi}(v_\pm) = m_\pm(m_\pm \mp 4{A^\prime}_\pm).
\ee
These expressions ensures that the solution connects critical points of the potential. Moreover, these points are minima of the potential whose values are negative for $AdS$ and null for Minkowski geometries. With the inclusion of the Cuscuton term in the Lagrange density, as in Eq.~\eqref{cuscuton}, the condition $V_\phi(v_\pm) =0$ is not ensured for constant $f(\phi)$ and ${A^\prime}_\pm \neq0$. To remedy this possibility, one takes advantage of the Galileon-like term in Eq.~\eqref{lagrange}, as we shall see below. 

For general $\alpha$ and $\beta$, considering the exponential falloff in Eq.~\eqref{phiasy}, we get from Eq.~\eqref{eqeinsteina} that the warp function behaves asymptotically $(y\to\pm\infty)$ as
\be
\begin{aligned}
	\!\!A^\prime &= {A^\prime}_\pm\mp\frac{2\kappa_\pm}{3}\left( 2\alpha m_\pm\pm\left(8\alpha{A^\prime}_\pm +\beta\right)\right)e^{- m_\pm |y|}\\
	\!\! &\pm\!\frac{\kappa_\pm^2}{9}\!\left(\left(3\!+\!32\alpha^2\right)m_\pm \!\pm\!16\alpha\left(8\alpha{A^\prime}_\pm \!+\!\beta\right)\right)\! e^{- 2m_\pm |y|}.
\end{aligned}
\ee
In contrast to the standard case, we now have a contribution of terms that engender exponential decay. The behavior of the potential can be found through Eq.~\eqref{eqeinsteinv}, which leads us to the following expressions
\bes\label{vpm}
\bal
& V(v_\pm) = -3{A^\prime}_\pm^2,\\ \label{vphigeral}
& V_\phi(v_\pm) = 4{A^\prime}_\pm(8\alpha{A^\prime}_\pm +\beta),\\
& V_{\phi\phi}(v_\pm) = \frac13\left(3+32{\alpha}^{2}\right) m_\pm  \left(m_\pm\mp4{A^\prime}_\pm \right) \nonumber\\
&\hspace{16mm}-\frac83\, \left(16\alpha{A^\prime}_\pm+\beta\right)  \left( 8\alpha{A^\prime}_\pm+\beta \right).
\eal
\ees

Let us first suppose that $v_\pm$ are critical points of the potential. This implies that $V_\phi(v_+)=0$ and $V_\phi(v_-)=0$, requiring the need of restrictions on the asymptotic behavior of the warp function, which can be found from Eq.~\eqref{vphigeral}. This also implies that one cannot obtain a solution connecting $AdS$ geometries. Inevitably, one of the tails of the solution has to connect a Minkowski geometry (${A^\prime}_+=0$ or ${A^\prime}_-=0$). We consider the situation where ${A^\prime}_-=0$, and ${A^\prime}_+=-\beta/8\alpha$ which gives us a $M$-$AdS$ brane, requiring the parameters $\alpha$ and $\beta$ to have the same sign.

We then look into each side of the brane, asymptotically. At the left tail, in which we have taken ${A^\prime}_-=0$, we have from Eq.~\eqref{vpm} that $V(v_-) = 0$,  $V_\phi(v_-)=0$ and
\be\label{vppmenos}
    V_{\phi\phi}(v_-) = \frac13\left(3+32{\alpha}^{2}\right) m_-^2 -\frac83\beta^2.
\ee
On the other hand, at the right tail, where we considered ${A^\prime}_+=-\beta/8\alpha$, we have the potential behaving as $V(v_+)=-3\beta^2/64\alpha^2$, $V_\phi(v_+)=0$, and
\be\label{vppmais}
    V_{\phi\phi}(v_+)=\frac13\left(3+32{\alpha}^{2}\right)  m_+\left(m_++\frac{\beta}{2\alpha} \right).
\ee
From the above expression, as $\alpha$ and $\beta$ must have the same sign, $v_+$ defines a minimum of the potential. Notwithstanding that, the same cannot be stated about $v_-$, which may not lead to a minimum of the potential, depending on the values of $\alpha$, $m_-$ and $\beta$. To ensure that it is a minimum, one must choose the parameters carefully.

As we have found the conditions to make the solution connect minima of the potential, we now investigate a procedure to reduce Eq.~\eqref{eqeinsteina} to first order. To do so, we introduce an auxiliary function $h=h(\phi)$ that obeys the equation
\be\label{foh}
\phi^\prime = h_\phi\, e^{-16\alpha\phi/3}.
\ee
The above expression can be used in Eq.~\eqref{eqeinsteina}, which can be integrated to give
\be\label{aprimemodel}
A^\prime\!=\!-\frac{\beta}{8\alpha}\!+\!\frac43\alpha \phi'\!-\! \frac29\left(3\!+\!32\alpha^2\right)\big(h\!+\!\tilde{h}_0\big)e^{-16\alpha\phi/3},
\ee
where $\tilde{h}_0$ is an integration constant. This, combined with Eq.~\eqref{eqeinsteinv}, allows us to find the explicit form of the potential as a function of the scalar field, $V(\phi)$.

To illustrate our procedure, we work with the model described by
\be\label{phiprime}
\phi^\prime = \lambda(1-\phi^2),
\ee
where $\lambda$ is a positive real parameter. This equation is solved by the function
\be\label{tanh}
\phi(y) = \tanh(\lambda y),
\ee
which connects the values $v_\pm = \pm1$ that define local minima of the potential. This solution engenders the same asymptotic behavior of Eq.~\eqref{phiasy}, with $\kappa_{\pm}=2$ and $m_{\pm}=2\lambda$. To calculate the auxiliary function $h(\phi)$ in this case, we can use Eqs.~\eqref{foh} and \eqref{phiprime} to get $h_\phi = \lambda e^{16\alpha\phi/3}(1-\phi^2)$, which can be integrated to give
\be
h(\phi) \!=\! \frac{3\lambda}{2048\alpha^3}\!\left(128\alpha^2\!\left(1-\phi^2\right)+48\alpha\phi-9\right)\!e^{16\alpha\phi/3}.
\ee
By using this in Eq.~\eqref{aprimemodel}, one can show that the warp function becomes
\be\label{alphi}
\begin{aligned}
	\!\!\!A^\prime=&\, \frac{9 \lambda\!-\!32 \alpha^2(4\beta\!+\!\lambda )}{1024 \alpha ^3}-\frac{\lambda}{64\alpha ^2}\! \left(3+32\alpha^2\right)\!\phi\\
	\!\!\!&+\frac{\lambda\phi^2}{8\alpha}-h_0 \,e^{-16 \alpha  \phi/3},
\end{aligned}
\ee
where we redefined the integration constant in a convenient form, as $h_0=2\left(3\!+\!32\alpha^2\right)\tilde{h}_0/9$. The above expression can be combined with Eq.~\eqref{phiprime} in Eq.~\eqref{eqeinsteinv} to calculate the potential, $V(\phi)$. Its expression is cumbersome, so we omit it here.

Asymptotically for $y\to\pm\infty$, the derivative of the warp function given above behaves as
\be\label{warp4}
	A^\prime_{\pm}= -\frac{\beta}{8\alpha} + \frac{\lambda\left(3+32\alpha^2\right)\left(3\mp 16\alpha\right)}{1024\alpha^3}-h_0 \,e^{\mp16\alpha/3}.
\ee
Let us first consider the case with $h_0=0$. For this choice, if we want to get a solution that goes from a Minkowski geometry at $y\to-\infty$ to an $AdS$ one at $y\to+\infty$, we need to impose that $A^\prime_-~=~0$, which leads us to  $\lambda = 128 \alpha ^2 \beta/\left(\left(3+16\alpha\right)\left(3+32 \alpha^2\right)\right)$, and also $A^{\prime}_+=-4\beta/\left(3+16\alpha\right)$. These conditions makes $V_\phi(v_-)=0$, as we have shown before, and
\be
\begin{aligned}
V_\phi(v_+)=\frac{16 (16 \alpha -3) \beta ^2}{(16 \alpha +3)^2}.
\end{aligned}
\ee
Note that, in order to have $V_\phi(v_+)=0$, we must choose $\alpha=3/16$. However, this restriction leads to
\be
\begin{aligned}
V_{\phi\phi}(v_-)=\frac{50 \beta ^2}{33}\qquad\mbox{and}\qquad V_{\phi\phi}(v_+)=-\frac{82\beta^2}{33}.
\end{aligned}
\ee
These expressions shows that $\phi=v_-$ is a minimum point and $\phi=v_+$ is a maximum point of the potential. Therefore, the case $h_0=0$ does not allow for the point $\phi=v_+$ being a minimum of the potential. 

To circumvent this issue, let us now consider the general case in which $h_0\neq 0$. We take $A'_-=0$ in Eq. \eqref{warp4} and obtain that the integration constant must be
\ben \label{h0}
\!\!h_0\!=\!\frac{e^{-16\alpha/3}\!\left(512\alpha^3\lambda\!-\!128 \alpha^2\beta\!+\!96\alpha^2\lambda\!+\!48 \alpha\lambda\!+\!9\lambda\right)}{1024\alpha^3}\!.
\een
As expected from the discussion above Eq.~\eqref{vppmenos}, it leads to $V(v_-)=V_{\phi}(v_-)=0$. To get a brane with an $AdS$ geometry at its right tail ($y\to+\infty$), we consider $A^\prime_+<0$ and $V_\phi(v_+)=0$, which can be attained by taking
\be\label{lambda}
\lambda = -\frac{64e^{-16\alpha/3}\alpha^2\beta}{\left(3+32\alpha^2\right) \left(3 \sinh \left(\frac{16\alpha}{3}\right)-16 \alpha\cosh\left(\frac{16\alpha}{3}\right)\right)}.
\ee
At $\phi=v_+$, the potential has the value $V(v_+)=-3\beta^2/64\alpha^2$, as expected from the expression right above Eq.~\eqref{vppmais}. This  is always negative, as it usually occurs for $AdS$ geometries. In this situation, from Eq.~\eqref{vppmais} we also have
\be
    V_{\phi\phi}(v_+)=\frac23\left(3+32{\alpha}^{2}\right)  \lambda\left(2\lambda+\frac{\beta}{2\alpha} \right),
\ee
with $\lambda$ as in Eq.~\eqref{lambda}. From this expression, one can show that $v_+=1$ is a minimum of the potential, matching with the discussion below Eq.~\eqref{vppmais} for solutions with exponential tails. On the other hand, we have
\be    V_{\phi\phi}(v_-) = \frac43\left(3+32{\alpha}^{2}\right)\! \lambda^2 -\frac83\beta^2,
\ee
with $\lambda$ as in Eq.~\eqref{lambda}, again. So, to ensure that $v_-=-1$ is a minimum of the potential, we impose the conditions $\alpha<0$ and $0<\alpha<\alpha_*$, where $\alpha_*=0.096$ is obtained numerically. When choosing $\alpha$ and $\beta$, one must remember that these parameters must have the same sign.

\begin{figure}[t!]
\centering
\includegraphics[width=6cm,trim={0cm 0cm 0 0cm},clip]{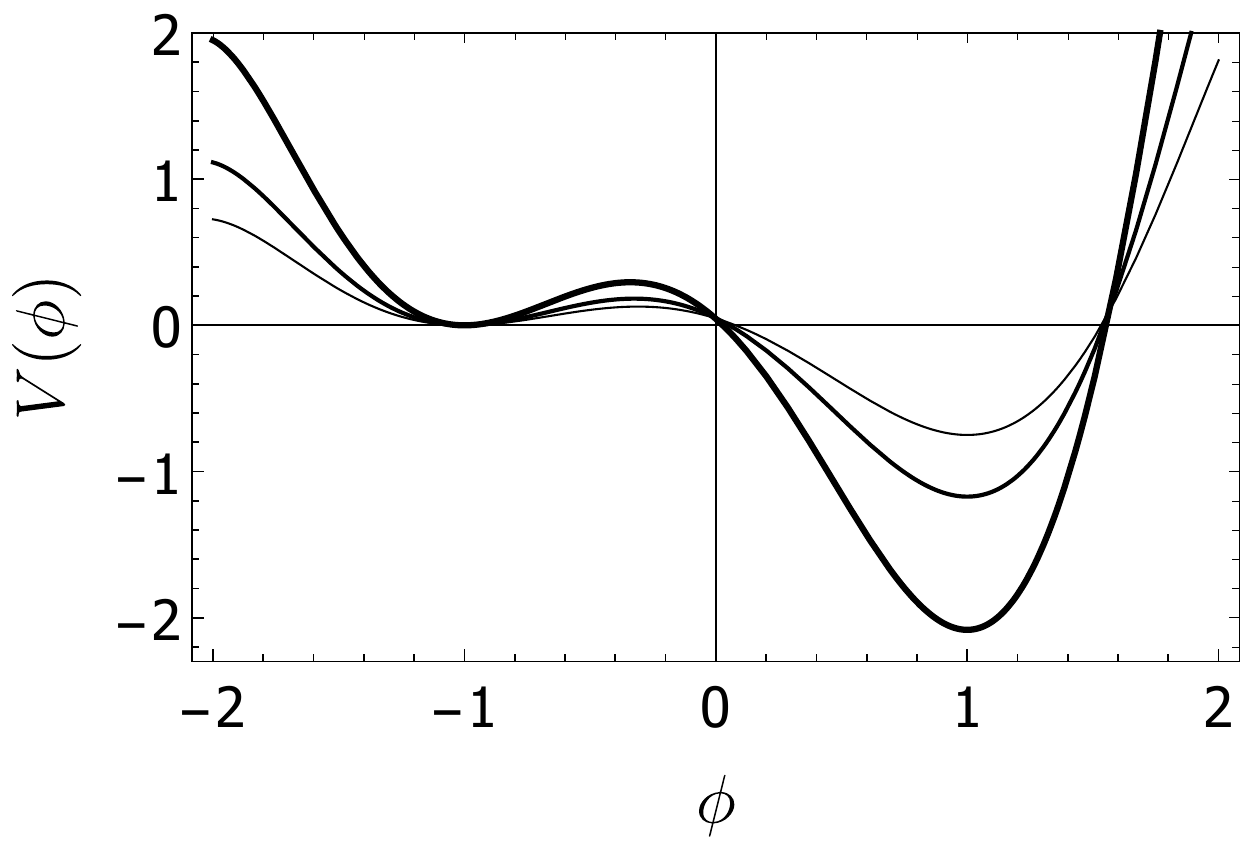}
\includegraphics[width=6cm,trim={0cm 0cm 0 0cm},clip]{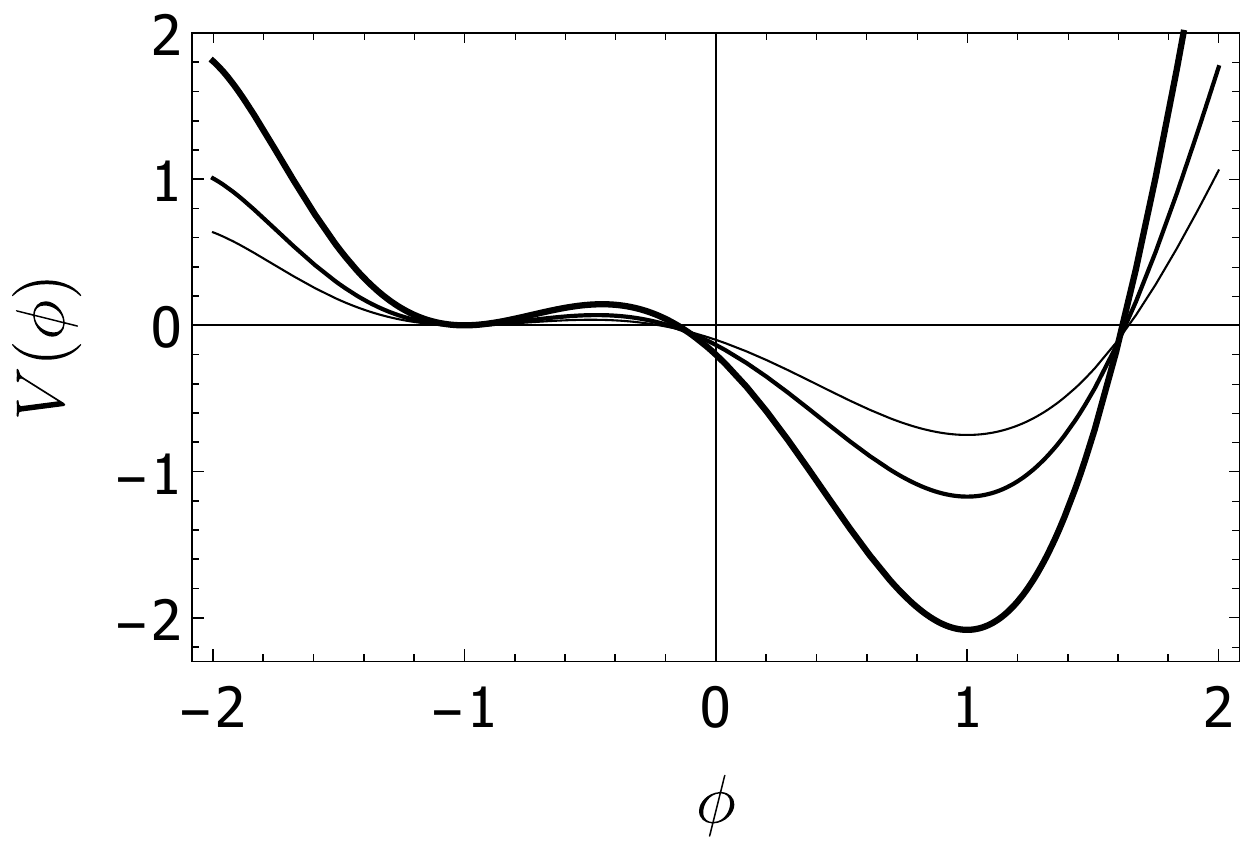}
\caption{The potential for the $M$-$AdS$ brane, for some values of $\beta$ and $\alpha$, with $h_0$ and $\lambda$ given by Eqs. \eqref{h0} and \eqref{lambda}. In the top panel, we show the potential for $\beta=-0.1$ and $\alpha = -0.025,-0.02$ and $-0.015$. In the bottom panel, we display the potential for $\beta=0.1$ and $\alpha = 0.015,0.02$ and $0.025$. In both panels, the width of the lines increases with $\alpha$.}
\label{figpot}
\end{figure}

Considering the conditions obtained above, we display, in Fig.~\ref{figpot}, the potential $V(\phi)$ in the interval $\phi\in[-2,2]$ for some values of $\beta$ and $\alpha$. Notice that the minimum at $v_-=-1$ is also a zero whilst the other one, at $v_+=1$, is not, attaining the value $V(v_+)=-3\beta^2/64\alpha^2$ as given above Eq.~\eqref{vppmais}. One can show that $V(\pm\infty) \to -\infty$, so it has maxima outside the range considered in this figure.

To calculate the warp function, we combine the solution in Eq.~\eqref{tanh} with Eq.~\eqref{alphi} to get
\be
\begin{aligned}
\!\!A^\prime=&\, \frac{9 \lambda\!-\!32 \alpha^2(4\beta\!+\!\lambda )}{1024 \alpha^3}\!-\!\frac{\lambda}{64\alpha^2}\! \left(3\!+\!32\alpha^2\right)\!\tanh(\lambda y)\\
\!\!&+\frac{\lambda}{8\alpha}\tanh^2(\lambda y)-h_0 \,e^{-16 \alpha  \tanh(\lambda y)/3},
\end{aligned}
\ee
where $h_0$ is given by Eq.~\eqref{h0} and $\lambda$ is as in Eq.~\eqref{lambda}. By integrating it, we get the expression
\be\label{warpfunction}
\begin{aligned}
A(y) &\!=\frac{3\!+\!32\alpha^2}{64\alpha^2}\!\ln (\sech(\lambda y))\!+\!\frac{\left(32\alpha^2  (3\lambda\!-\!4\beta)\!+\!9\lambda\right)\!y}{1024\alpha^3}\\
&-\!\frac{\tanh(\lambda y)}{8\alpha} \!+\!\frac{h_0}{2\lambda}\!\left(\!e^{-\frac{16 \alpha }{3}}\text{Ei}\!\left(\xi_-\right)\!-\!e^{\frac{16 \alpha }{3}} \text{Ei}\!\left(\xi_+\right)\!\right)\\
&-\frac{h_0}{2 \lambda}\left(\!e^{-\frac{16\alpha}{3}}\text{Ei}\!\left(16\alpha/3\right)\!-\!e^{\frac{16\alpha}{3}} \text{Ei}\!\left(-16\alpha/3\right)\!\right)\,,
\end{aligned}
\ee
where $\xi_{\pm}=-(16\alpha/3)(\tanh (\lambda  y)\pm 1)$ and $\rm Ei$ denotes the Exponential Integral function. Here we have used that $A(0)=0$.

\begin{figure}[t!]
\centering
\includegraphics[width=6cm,trim={0cm 0cm 0 0cm},clip]{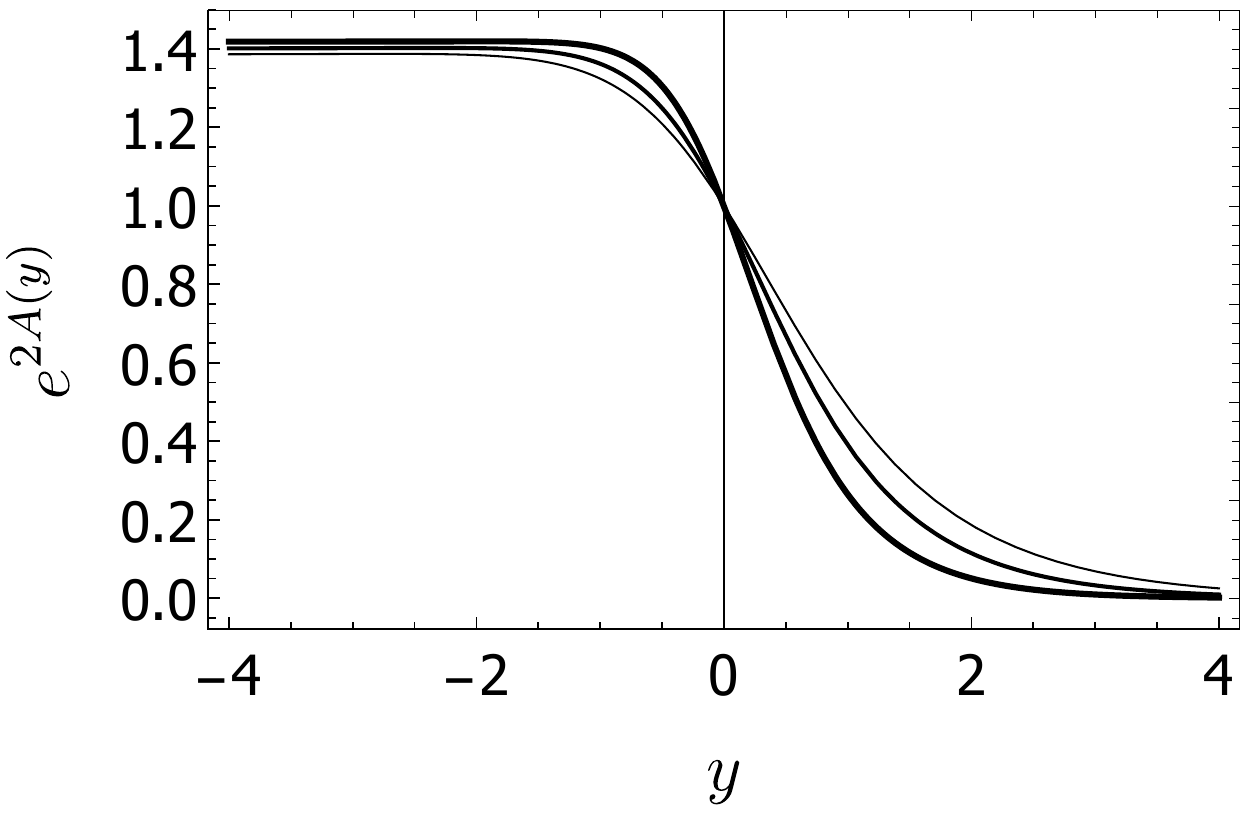}
\includegraphics[width=6cm,trim={0cm 0cm 0 0cm},clip]{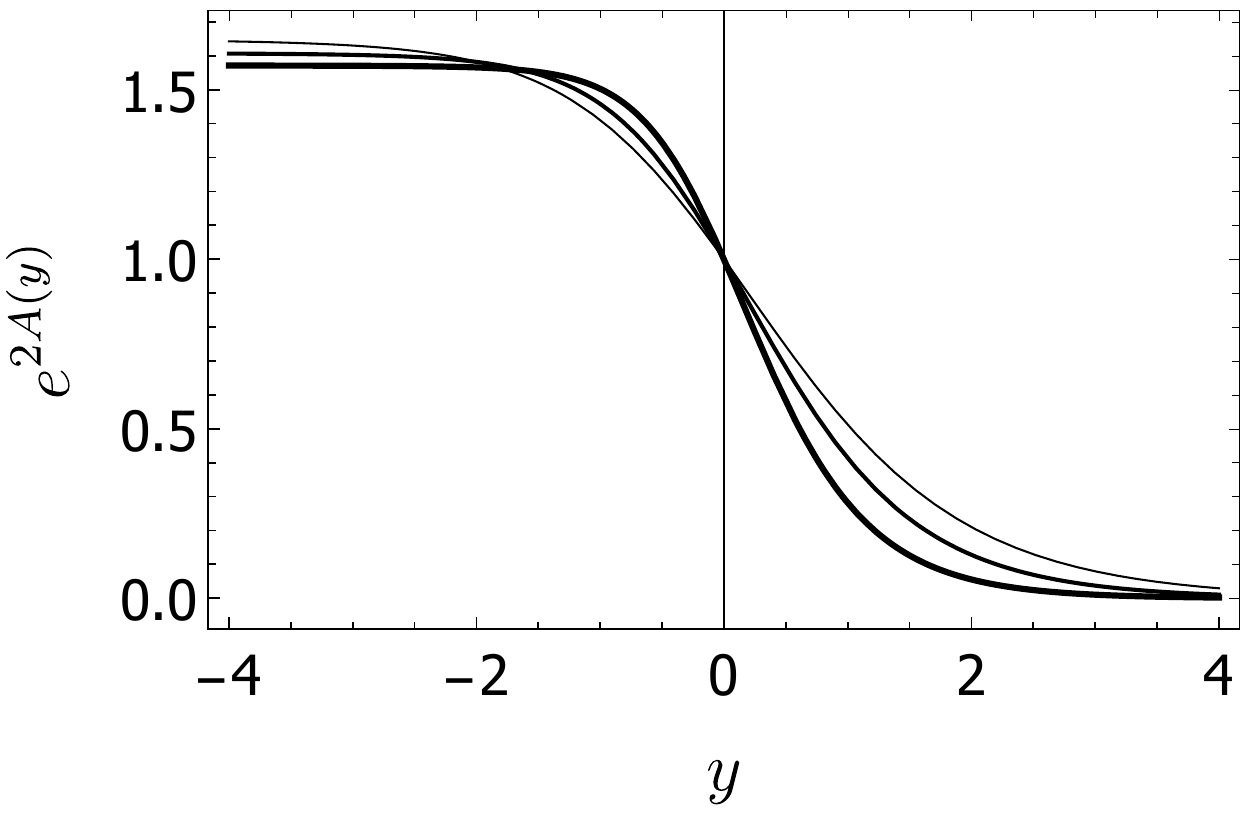}
\caption{The warp factor $e^{2A(y)}$ associated to the warp function in Eq.~\eqref{warpfunction}, depicted with the same parameters of Fig. \ref{figpot}.}
\label{figwfactor}
\end{figure}

In Fig.~\ref{figwfactor}, we display the warp factor $e^{2A(y)}$ for the warp function in Eq.~\eqref{warpfunction} for some values of $\beta$ and $\alpha$. We see that $e^{2A(-\infty)}>0$ and $e^{2A(\infty)}=0$, showing that our solution describes a $M$-$AdS$ brane.  

\begin{figure}[t!]
\centering
\includegraphics[width=6cm,trim={0cm 0cm 0 0cm},clip]{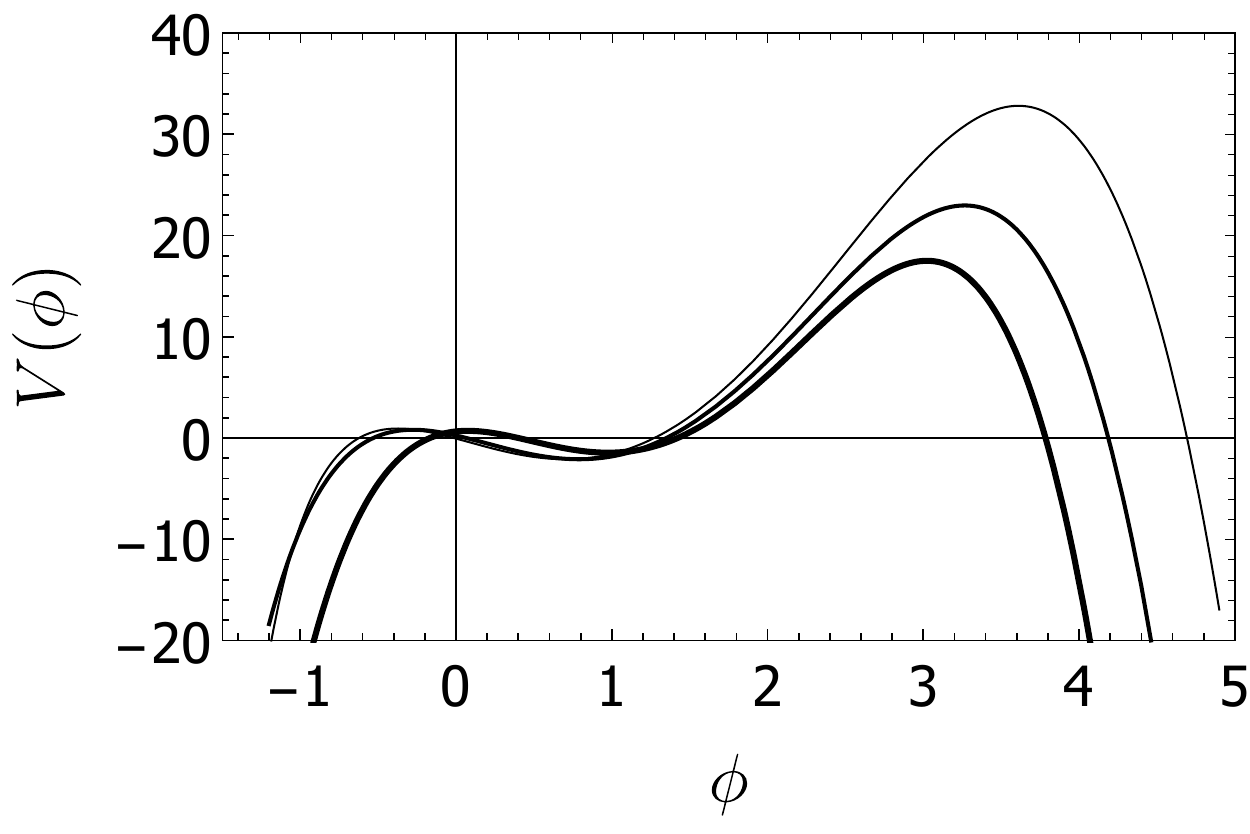}
\includegraphics[width=6cm,trim={0cm 0cm 0 0cm},clip]{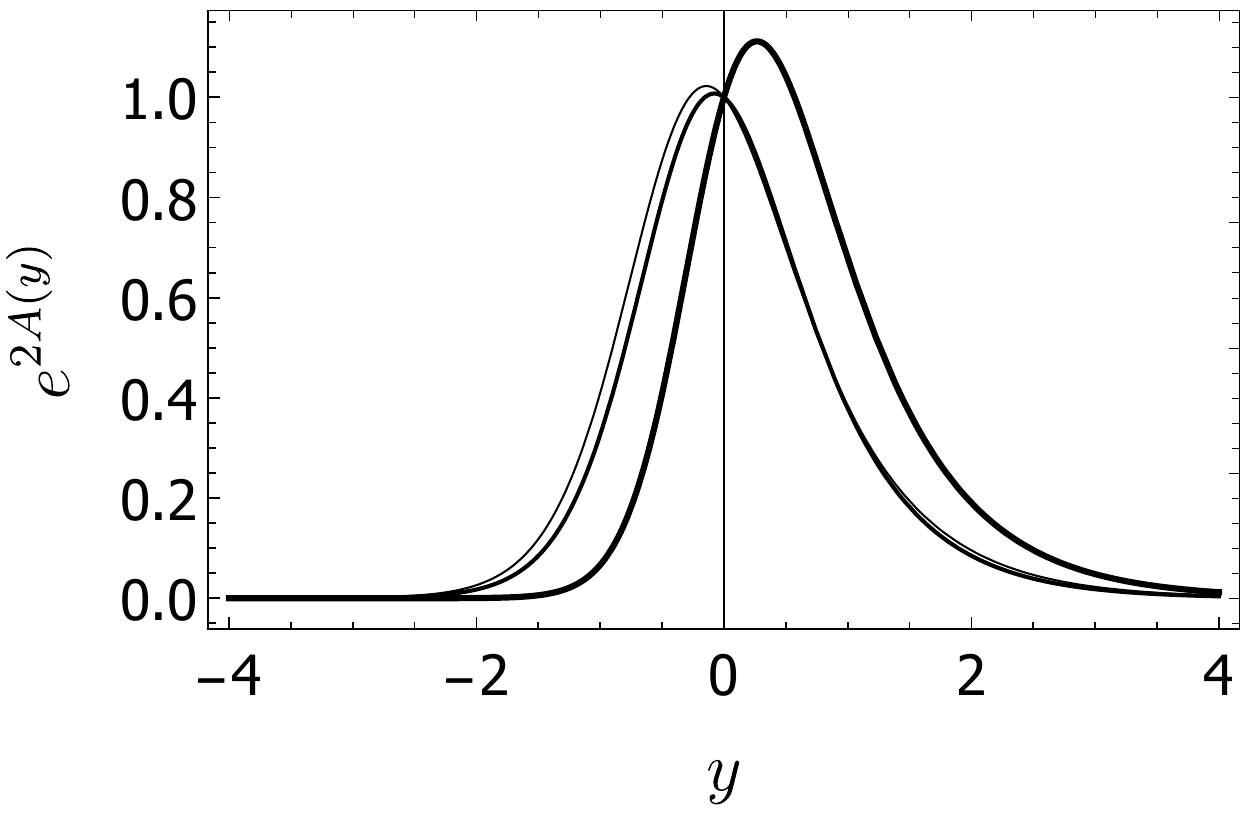}
\caption{In the top panel we display the potential $V(\phi)$ for the $AdS$-$AdS$ brane, with $\beta=\lambda=1$, $h_0=-0.1$ and $\alpha = 0.2, 0.3$ and $0.4$. In the bottom panel, we depict its warp factor associated to the warp function in Eq.~\eqref{warpfunction} for the same values of parameters. In both panels, the width of the lines increases with $\alpha$. }
\label{figwfactorAds}
\end{figure}

We can further study the system, to investigate the situation where the warp factor connects two $AdS$ geometries, with $e^{2A(\pm\infty)}=0$. This is the $AdS$-$AdS$ brane, and it can be implemented with the same Eq. \eqref{warpfunction}, but now the constraints in the parameters $h_0$ and $\lambda$ and in Eqs. \eqref{h0} and \eqref{lambda} have to be discarded. 
The results are depicted in Fig.~\ref{figwfactorAds}, showing the potential and warp factor for some values of $\alpha$, $\beta$, $\lambda$ and $h_0$. In this case, the scalar field solution does not connect minima of the potential and the brane is asymmetric, due to the presence of the Cuscuta-Galileon dynamics.

\begin{figure}[t!]
\centering
\includegraphics[width=6cm,trim={0cm 0cm 0 0cm},clip]{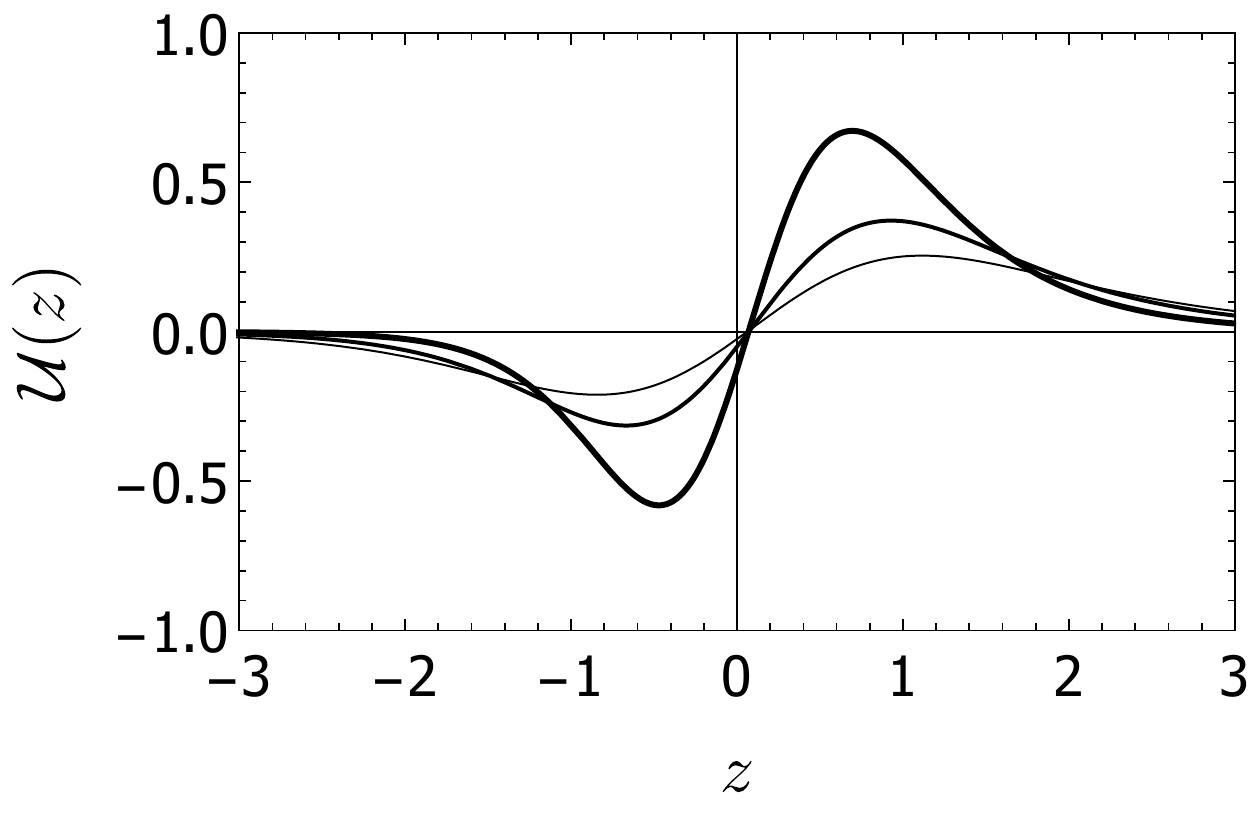}
\includegraphics[width=6cm,trim={0cm 0cm 0 0cm},clip]{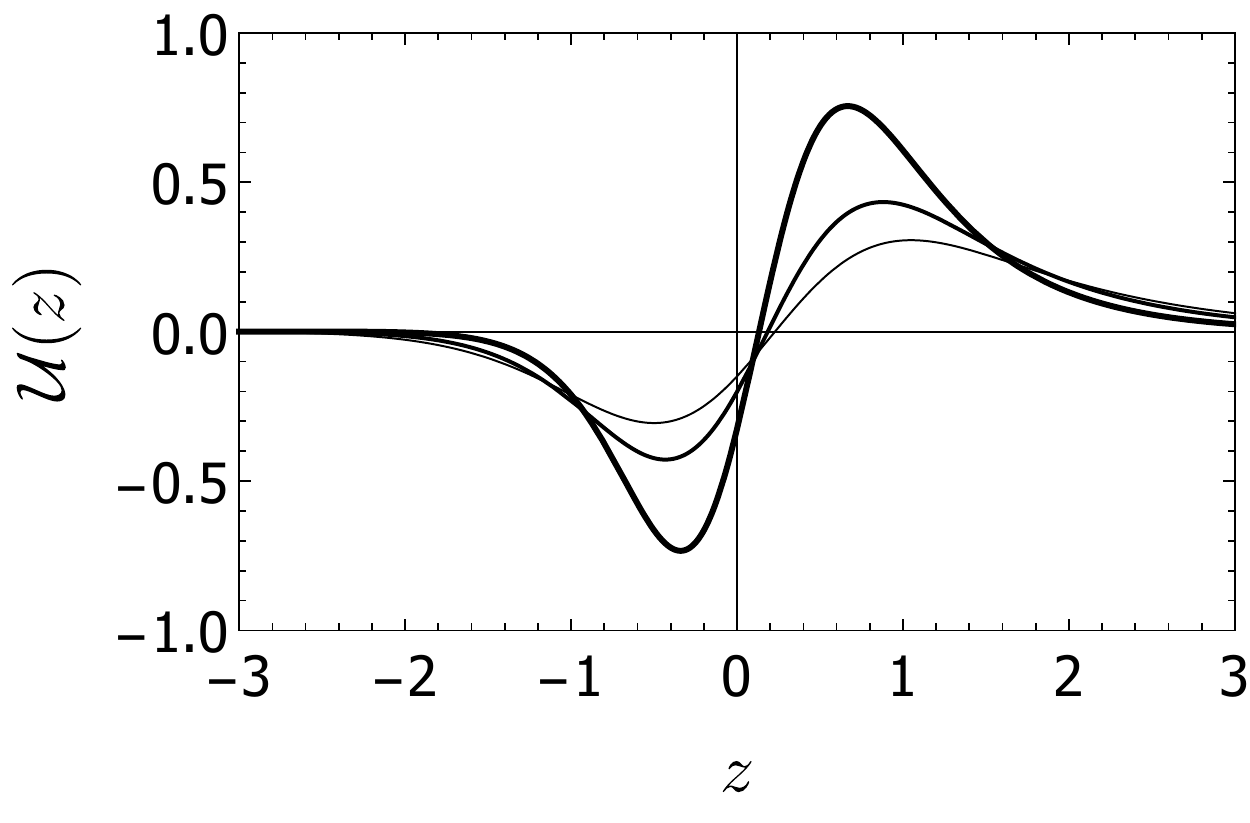}
\caption{The stability potential for the warp factor of the $M$-$AdS$ brane, depicted with the same parameters of Fig. \ref{figpot}.}
\label{figstab}
\end{figure}

\begin{figure}[t!]
\centering
\includegraphics[width=6cm,trim={0cm 0cm 0 0cm},clip]{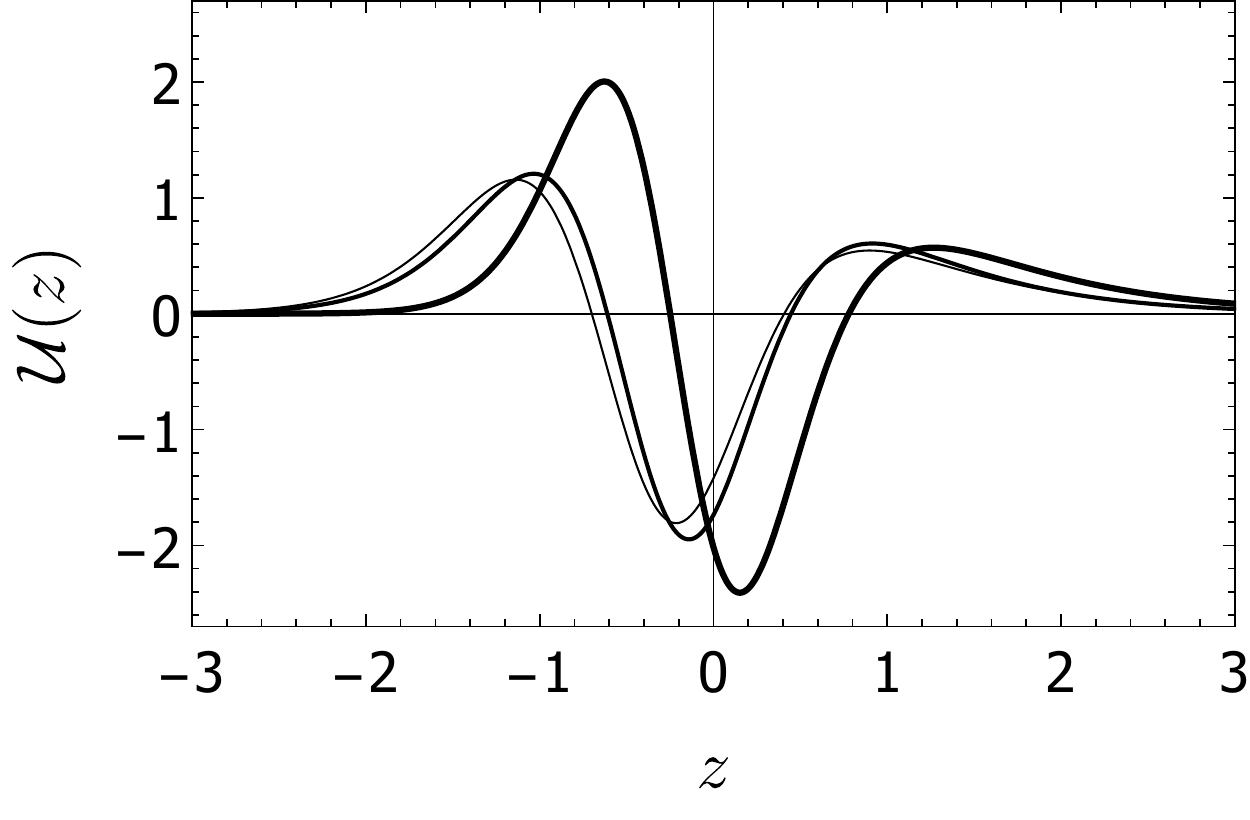}
\caption{The stability potential for the warp factor of the $AdS$-$AdS$ brane, depicted with the same parameters of Fig. \ref{figwfactorAds}.}
\label{figstab2}
\end{figure}

Let us now investigate linear stability of the gravity sector of the brane considering
\ben
ds^2=e^{2A(y)}\big(\eta_{\mu\nu}+h_{\mu\nu}(x^\mu,y)\big)dx^\mu dx^\nu-dy^2,
\een
in which $h_{\mu\nu}$ denotes the small fluctuations, where we have followed the lines of Ref.~\cite{mannheim} to take the axial gauge, $h_{a4} = 0$, that is convenient for the study and depends on the extra dimension and on the 4-dimensional vector $x^\mu$. We also take small perturbations around the scalar field solution, in the form $\phi\to\phi(y)+\xi(x^\mu,y)$, where $\phi(y)$ denotes the static solution of Eqs.~\eqref{eqeinstein}. These fluctuations can be inserted into Einstein's equation \eqref{einstein}, which leads us to three non null components labeled by $G_{\mu4}^{(1)}=2T_{\mu4}^{(1)}$, $G_{44}^{(1)}=2T_{44}^{(1)}$ and $G_{\mu\nu}^{(1)}=2T_{\mu\nu}^{(1)}$, where $G^{(1)}_{ab}$ is the term of the Einstein tensor that contains the fluctuations at first order. In particular, the latter component leads us to
\ben
\begin{aligned}
 &h_{\mu\nu}''+4A'h_{\mu\nu}'-e^{-2A}\Box h_{\mu\nu}\\
&+e^{-2A}\big(\partial_\mu\partial^\gamma h_{\gamma\nu}+\partial_\nu\partial^\gamma h_{\gamma\mu}-\partial_\mu\partial_\nu h\big)\\
&=\eta_{\mu\nu}\Big(h''+e^{-2A}\big(\partial^\gamma \partial^\sigma h_{\gamma\sigma}-\Box h\big)\\
&+4A'h'+4V_\phi \xi+4\phi'\xi'-8\alpha\xi''\Big)\,,
\end{aligned}
\een
where we have defined $h=\eta^{\mu\nu}h_{\mu\nu}$. We then proceed as usual and take the transverse and traceless (TT) condition for $h_{\mu\nu}$, i.e. $\partial^\mu h_{\mu\nu}=0$ and $h=0$. This makes all the components vanish, such that only the one in the above equation survives. Also, we take $dy=e^{A}dz$ and $h_{\mu\nu}(x^\mu,z)=e^{i \omega_\mu x^\mu}e^{-3A(z)/2}H_{\mu\nu}(z)$ to get the following  Schr\"odinger-like eigenvalue equation
\ben\label{eqpertu}
\Big(-\frac{d^2}{dz^2}+U(z)\Big)H_{\mu\nu}=\omega^2H_{\mu\nu},
\een
where the stability potential is given exclusively in terms of derivatives of the warp function, in the form
\begin{equation}\label{eq12}
    U(z)=\frac{9}{4} A_z^2+\frac{3}{2} A_{zz}.
\end{equation}
In Fig.~\ref{figstab}, we display the above stability potential when the warp factor connects $M$ and $AdS$ geometries, for some values of $\alpha$ and $\beta$. In Fig.~\ref{figstab2}, we display the stability potential for the case where the warp factor connects two $AdS$ geometries, using $\beta=\lambda=1$, $h_0=-0.1$ and $\alpha = 0.2, 0.3$ and $0.4$.
One can show that the operator in the left hand side of Eq.~\eqref{eqpertu} can be written as the product operators,  ${\cal S}^{\dagger}{\cal S}\,H_{\mu\nu}=\omega^2\,H_{\mu\nu}$, where ${\cal S}=-d/dz-3A_z/2 $ and ${\cal S}^{\dagger}=d/dz-3A_z/2$. This factorization ensures the linear stability of the gravity sector of the brane, as the operator associated to the eigenvalue equation \eqref{eqpertu} is non negative, implying that $\omega^2\geq0$. The localization of gravity in the braneworld scenario was previously investigated in several works, in particular, in  Refs.~\cite{Randall:1999vf,as7,dgp}. For solutions of the $AdS$-$AdS$ brane, the 4-dimensional Newtonian potential is given by the sum of contributions of the zero mode and the correction term associated to the exchange of Kaluza-Klein modes \cite{Randall:1999vf}. In the case in which one has the interpolation of $M$-$AdS$ geometries, the Newtonian potential arises due to the correction term alone. In both cases, the correct 4-dimensional Newtonian behavior can be recovered at short distances \cite{dgp,as7}.

In summary, we have investigated the Cuscuta-Galileon model described by the Lagrange density in Eq.~\eqref{lagrange}. The Cuscuton term in Eq.~\eqref{cuscuton} engender a function which only depends on the scalar field, $f(\phi)$. If $f(\phi)$ is constant, the model \eqref{cuscuton} cannot support scalar field solutions connecting minima of the potential $V(\phi)$. However, after adding the Galileon term in Eq.~\eqref{lagrange}, we have shown that it conspires against the Cuscuton term in the Einstein's equations, leaving room to construct interesting solutions that connect minima of the potential. Among the results, we recall that we were interested in having the minima connected by the solutions, so we have first investigated the behavior of the brane asymptotically, with the scalar field having exponential tails, as in Eq.~\eqref{phiasy}. This led us to find conditions for the parameters $\alpha$, $\beta$ and $m_-$, which must be chosen carefully. In particular, $\alpha$ and $\beta$ must have the same sign.  
Moreover, since the investigation involves nonlinear differential equations of second order, we have introduced an auxiliary function $h(\phi)$ that allows for the presence of first order equations compatible with the Einstein's ones. To illustrate our procedure, we have taken the well-known hyperbolic tangent profile in Eq.~\eqref{tanh} and found the warp factor associated to the brane that interpolates a Minkowski ($M$) and an Anti de-Sitter ($AdS$) geometry. We have also investigated the stability of the gravity sector of the brane and shown that the operator that governs the eigenvalue equation can be factorized, ensuring linear stability.

There are several perspectives, in particular, the case concerning asymmetry of the brane and the cosmic acceleration \cite{CA1,CA2}. The Cuscuta-Galileon model \eqref{lagrange} can also be investigated in other scenarios of generalized theories of gravity \cite{p1,p2,p3,p4,p5,p6}. The study of time-dependent configurations in the five-dimensional spacetime is also of interest, as it can shed light on the number of degrees of freedom that propagates in the theory \cite{dof1,dof2}. Another possibility concerns absence of gravity, in the case of flat spacetime, in the study of kinks \cite{vachaspati}, vortices \cite{novortex} and monopoles \cite{thooft,polyakov}. Some of these issues are currently being considered and shall be reported elsewhere.

This work is supported by the Brazilian agencies Conselho Nacional de Desenvolvimento Cient\'ifico e Tecnol\'ogico (CNPq), grants No. 303469/2019-6 (DB) and No. 310994/2021-7 (RM), Paraiba State Research Foundation (FAPESQ-PB), grants No. 0003/2019 (RM) and No. 0015/2019 (DB and MAM), and Federal University of Para\'iba (UFPB/PROPESQ/PRPG) project code PII13363-2020.


\end{document}